%% file: main.tex
\documentclass{article}

\usepackage{arxiv}

\usepackage[utf8]{inputenc} 
\usepackage[T1]{fontenc}    
\usepackage{hyperref}       
\usepackage{url}            
\usepackage{booktabs}       
\usepackage{amsfonts}       
\usepackage{nicefrac}       
\usepackage{microtype}      
\usepackage{lipsum}		
\usepackage{graphicx}
\usepackage{natbib}
\usepackage{doi}
\usepackage{orcidlink}
\usepackage{longtable}
\usepackage{makecell}
\usepackage{float}
\usepackage{amsmath}
\usepackage{subcaption}
\usepackage{caption}
\title{Beyond Current Boundaries: Integrating Deep Learning and AlphaFold for Enhanced Protein Structure Prediction from Low-Resolution Cryo-EM Maps}

\author{ \href{https://orcid.org/0009-0004-0136-2799}{\includegraphics[scale=0.06]{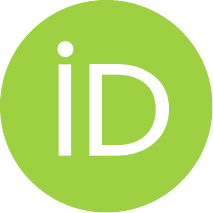}\hspace{1mm}Xin (Chloe) Ma} \\
	Division of Computing and Software Systems\\
	University of Washington Bothell\\
	Bothell, WA 98011 \\
	\And
	\href{https://orcid.org/0000-0001-7039-2589}{\includegraphics[scale=0.06]{orcid.pdf}\hspace{1mm}Dong Si} \\
	Division of Computing and Software Systems\\
	University of Washington Bothell\\
	Bothell, WA 98011 \\
}

\hypersetup{
pdftitle={Beyond Current Boundaries: Integrating Deep Learning and AlphaFold for Enhanced Protein Structure Prediction from Low-Resolution Cryo-EM Maps},
pdfauthor={Xin (Chloe) Ma, Dong Si},
pdfkeywords={cryo-EM, deep learning, AlphaFold, computational structural biology },
}

\begin{document}
\maketitle

\input{tex/abstract}

\input{tex/introduction}
\input{tex/related}

\input{tex/method}
\input{tex/results}
\input{tex/future}
\input{tex/conclusion}

\bibliographystyle{unsrtnat}
\bibliography{cas-refs}  






\end{document}

%% file: tex/abstract.tex
\begin{abstract}
Constructing atomic models from cryo-electron microscopy (cryo-EM) maps is a crucial yet intricate task in structural biology. While advancements in deep learning, such as convolutional neural networks (CNNs) and graph neural networks (GNNs), have spurred the development of sophisticated map-to-model tools like DeepTracer and ModelAngelo, their efficacy notably diminishes with low-resolution maps beyond 4 Å. To address this shortfall, our research introduces DeepTracer-LowResEnhance, an innovative framework that synergizes a deep learning-enhanced map refinement technique with the power of AlphaFold. This methodology is designed to markedly improve the construction of models from low-resolution cryo-EM maps. DeepTracer-LowResEnhance was rigorously tested on a set of 37 protein cryo-EM maps, with resolutions ranging between 2.5 to 8.4 Å, including 22 maps with resolutions lower than 4 Å. The outcomes were compelling, demonstrating that 95.5\% of the low-resolution maps exhibited a significant uptick in the count of total predicted residues. This denotes a pronounced improvement in atomic model building for low-resolution maps. Additionally, a comparative analysis alongside Phenix's auto-sharpening functionality delineates DeepTracer-LowResEnhance's superior capability in rendering more detailed and precise atomic models, thereby pushing the boundaries of current computational structural biology methodologies.
\end{abstract}

%% file: tex/introduction.tex
\section{Introduction}
Protein structure prediction involves predicting the three-dimensional (3D) structure of a protein from its amino acid sequence, which constitutes a critical component in computational biology. The capacity for precise delineation of protein configurations holds substantial promise for the advancement of life sciences and medicinal research \cite{deepmind2020}. Such capabilities would significantly expedite the elucidation of cellular building blocks and foster the development of advanced pharmacological interventions. Currently, the identification of new protein sequences is growing much faster than its corresponding solved experimental 3D structure \cite{dnastar2020}, so the exploration of alternative methods for protein structure prediction is becoming increasingly imperative. In this context, protein structure model prediction emerges as a critical mechanism to bridge the gap between the protein sequence information and its corresponding solved 3D structure, representing a foremost challenge within computational biology and chemistry. This produces a concerted focus on the development and refinement of computational algorithms and modeling techniques, which are essential in advancing our understanding and capabilities in this domain.

There are two main types of protein structure prediction methods: sequence-to-model and map-to-model. Sequence-to-model involves predicting a protein’s 3D structure directly from its amino acid sequence. One of the most successful examples of this approach is AlphaFold, an AI system developed by DeepMind. AlphaFold predicts a protein’s 3D structure from its amino acid sequence and regularly achieves accuracy competitive with experiments. It has been used to make over 200 million protein structure predictions, which are freely available to the scientific community through the AlphaFold Protein Structure Database \cite{alphafold}. 

\begin{figure}[h]
  \centering
  \includegraphics[width=0.7\linewidth]{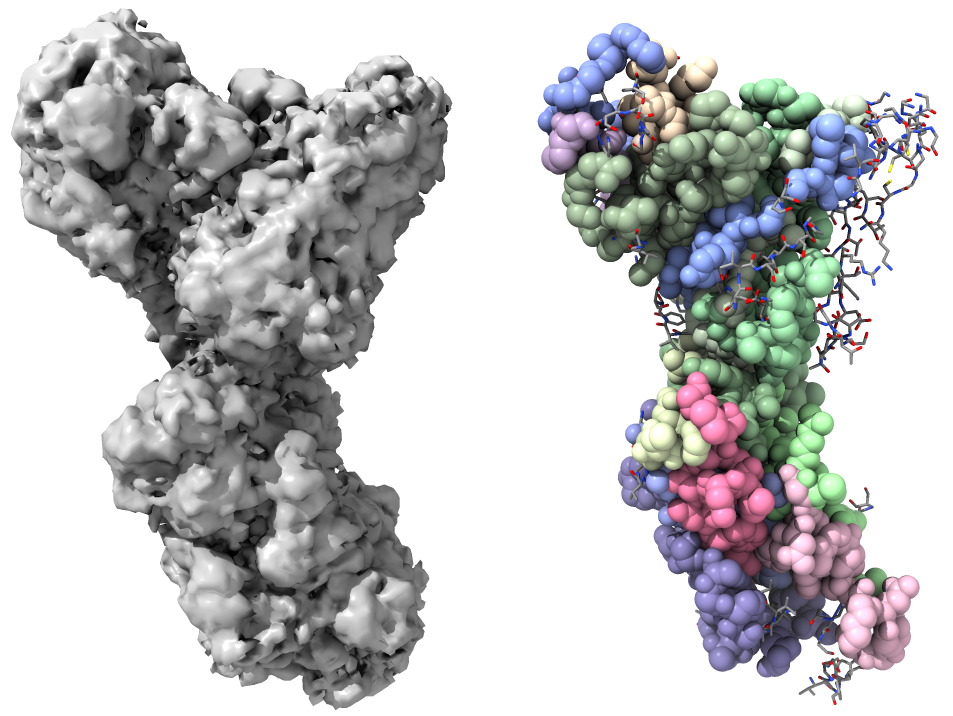}
  \caption{Utilizing high-resolution cryo-EM electron density maps to predict protein structure. Left: cryo-EM density map (EMD-15271). Right: Predicted all-atom structure by running ModelAngelo \cite{ModelAngelo}.}
  \label{fig:intro_example}
\end{figure}

\begin{figure}[h]
    \centering
    \begin{subfigure}[t]{0.48\textwidth}
        \centering
        \includegraphics[width=\linewidth]{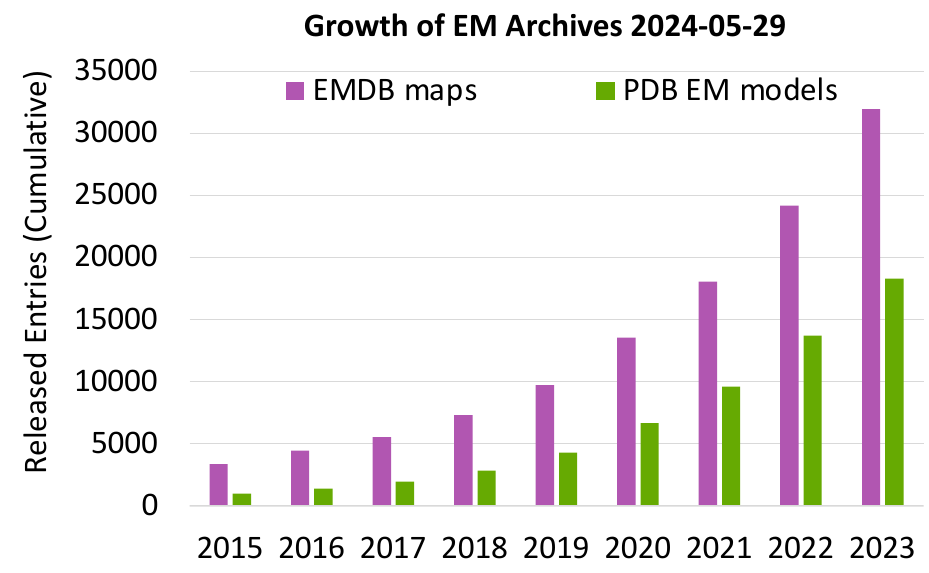}
        \caption{Growth of cryo-EM maps in Electron Microscopy Data Bank (EMDB) and solved protein structure models in Protein Data Bank (PDB) along with near-atomic EM technology improvement from 2015 to 2024. The gap between the number of cryo-EM maps and the number of solved models has widened each year, indicating an increasing number of cryo-EM maps awaiting structural modeling.}
    \label{fig:emdatabase1}
    \end{subfigure}%
    ~ 
    \begin{subfigure}[t]{0.48\textwidth}
        \centering
        \includegraphics[width=\linewidth]{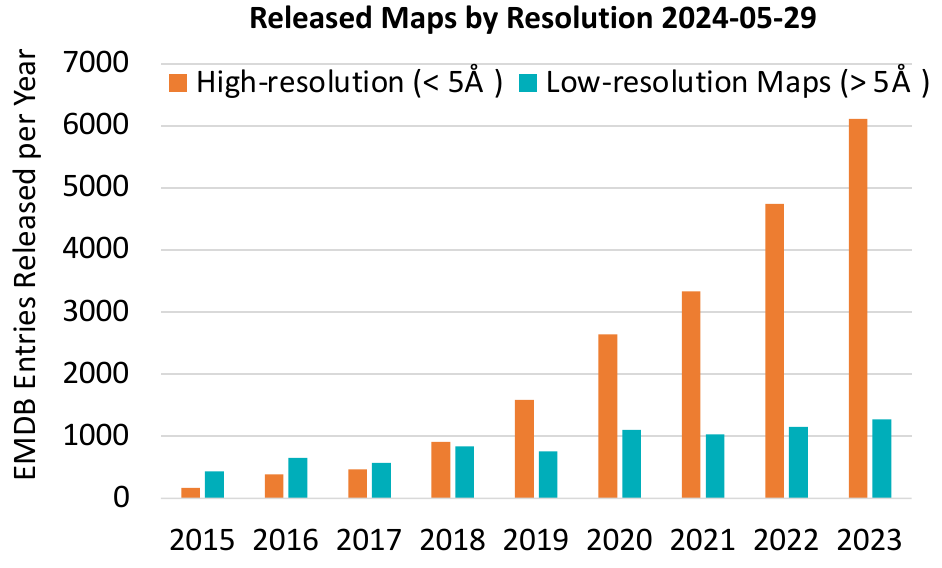}
    \caption{Released cryo-EM maps in the Electron Microscopy Data Bank (EMDB) categorized by resolution and year. A higher angstrom value indicates lower resolution. The number of high-resolution maps (resolution < 5 Å) has grown rapidly from 2015 to 2023, while a substantial number of low-resolution maps (resolution > 5 Å) remain available.}
    \label{fig:emdatabase2}
    \end{subfigure}
    \caption{Trend of EMDB maps and PDB EM models along the years. (Accessed on May 29th, 2024. \cite{emdb, pdb})}
\end{figure}

The other main method, map-to-model, involves using experimental data, such as electron density maps obtained from X-ray crystallography or cryo-electron microscopy (cryo-EM), to build a model of the protein structure. Fig. \ref{fig:intro_example} shows an example of utilizing cryo-EM map (left) to construct an all-atom model (right) for SARS Cov2 Spike RBD in a complex with Fab47 protein. The experimental data provides a “map” that can be used to guide the placement of atoms in the protein model \cite{Jisna2021}. Therefore, the quality of the map, quantified by its resolution in Angstroms (Å), plays an important role in determining the accuracy of the resulting predictive model. A lower value in Angstroms (Å) indicates a higher resolution in the cryo-EM map, which captures finer details and directly correlates with enhanced accuracy and structural integrity of the predicted 3D model. The resolution revolution in cryo-EM has not only greatly accelerated the experimental determination of protein structures but also established it as a powerful imaging technology \cite{kuehlbrandt2014resolution, liu2017cryoem}. The data presented in Fig. \ref{fig:emdatabase1} illustrates a positive correlation between the growing number of Electron Microscopy Data Bank (EMDB) \cite{emdb} maps and the increasing number of resolved ground-truth models in the Protein Data Bank (PDB) \cite{pdb}. However, a disparity remains between the increasing number of EMDB maps and the relatively fewer solved 3D protein models in PDB, indicating many maps still await modeling.

Recent innovations in map-to-model techniques have played a critical role in bridging this divide. For example, deep-learning-based protein structure prediction tools are now adept at producing high-quality models from cryo-EM maps finer than 5 Å. These improvements are reflected in the trends illustrated in Fig. \ref{fig:emdatabase1} and Fig. \ref{fig:emdatabase2}, where the increase in high-resolution EMDB maps has resulted in a corresponding rise in the number of PDB models \cite{EMDB_fig2}. A detailed examination of Fig. \ref{fig:emdatabase2} reveals that this upsurge in EMDB maps is primarily attributed to the availability of high-resolution maps (2 - 5 Å), while a substantial number of low-resolution maps are yet to be fully utilized. With increasing resolution, critical protein features like secondary structures and amino acid side chains become more discernible \cite{Rosenthal_2019}, underscoring the complexity and time-intensive nature of enhancing map resolution. Therefore, the development of automated tools capable of harnessing the potential of low-resolution maps is an urgent and vital task in advancing the field of map-to-model prediction.

Our research focuses on developing DeepTracer-LowResEnhance, a new tool that targets extending DeepTracer's ability to process low-resolution maps by using sequence information to integrate the strengths and overcome the limitations of existing models. It is designed to serve a diverse audience, ranging from academic researchers to medical practitioners, all engaged in the elucidation of complex three-dimensional biomolecular architectures.

%% file: tex/related.tex
\section{Related Works}
 
The field of protein structure prediction has seen substantial advancements, particularly with the advent of sophisticated computational methods. This chapter provides an overview of the key methodologies and techniques that have shaped the current landscape. Section 2.1 delves into map-to-model methods for protein structure prediction, including various prediction techniques and approaches for enhancing cryo-EM maps. Section 2.2 explores sequence-assisted methods, focusing on sequence-to-model techniques and hybrid approaches that combine multiple strategies. Finally, Section 2.3 addresses the limitations of current methods, emphasizing the challenges posed by low-resolution cryo-EM maps. Despite the proficiency of existing techniques in handling high-resolution data, there is a critical need for automated tools capable of leveraging low-resolution maps to advance the field of protein structure prediction.

\subsection{Map-to-model Methods for Protein Structure Prediction}
\subsubsection{Prediction Methods}
Accurate prediction of protein structures is significantly important in recent research. Particularly, 3D convolutional neural networks (CNNs) have driven the development of sophisticated all-atom protein structure modeling tools like DeepTracer \cite{Pfab_Phan_Si_2020}, DeepMM \cite{he2021full}, CR-I-TASSER \cite{zhang2022cr}, and Emap2sec \cite{subraman2022deep}. Also, the emergence of graph neural networks (GNNs)-based models like ModelAngelo \cite{ModelAngelo} has demonstrated exceptional performance and accuracy for high-resolution maps. However, existing map-to-model methods often struggle with low-resolution maps. In this section, we explore seminal works that have paved the way for the methodologies and techniques central to our study.

\textbf{DeepTracer}: Embodying the integration of deep convolutional neural networks (CNNs) and U-Net architectures, DeepTracer is a seminal tool in the field of protein structure determination. Engineered for efficiently delineating structures of multi-chain protein complexes, it predominantly processes high-resolution cryo-EM maps. As an evolution from DeepTracer 1.0, DeepTracer introduces the capability to predict DNA and RNA structures \cite{Nakamura_Meng_Zhao_Wang_Hou_Cao_Si_2023}, reflecting significant advancements in structural biology. 

\begin{figure}[h]
  \centering
  \includegraphics[width=0.7\linewidth]{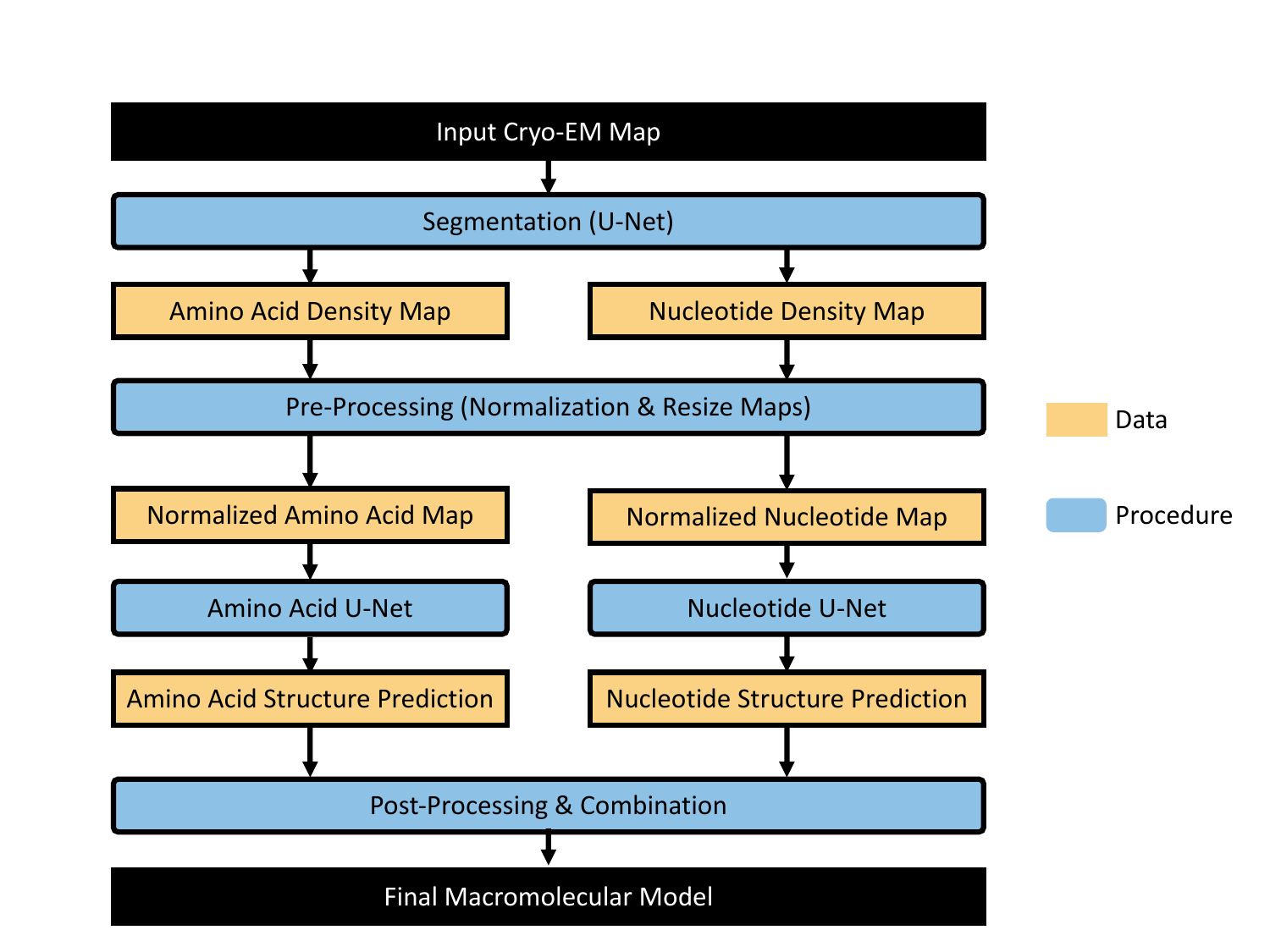}
  \caption{DeepTracer pipeline for analyzing cryo-EM maps to predict protein and nucleic acid structures. The process starts with the cryo-EM map input, followed by segmentation using a U-Net convolutional neural network (CNN) and normalization with resizing. A specialized nucleotide U-Net analyzes nucleotides, distinct from the amino acid U-Net. The final output combines the predicted amino acid structures with post-processed nucleotides.}
  \label{fig:DT_pipeline}
\end{figure}

The DeepTracer pipeline is a sophisticated process for analyzing cryo-EM maps to predict the structures of proteins and nucleic acids. As illustrated in Fig. \ref{fig:DT_pipeline}, the pipeline's initial step involves the input of cryo-EM maps. These maps undergo a crucial segmentation phase, utilizing a CNN with a sophisticated U-Net architecture. In the pre-processing stage, DeepTracer meticulously separates, normalizes, and resizes macromolecular densities present in the cryo-EM map. This process standardizes them to a cube size of $64^3$ $\text{Å}^3$, preparing them effectively for analysis by the nucleotide U-Net \cite{Pfab_Phan_Si_2020}. This nucleotide U-Net is distinct from the one used for amino acids, reflecting the different molecular structures of nucleotides and amino acids. During post-processing for refining the predicted structure, the Brickworx model plays a key role in finalizing the nucleotide structure, identifying double-stranded helical motifs in the cell \cite{Nakamura_Meng_Zhao_Wang_Hou_Cao_Si_2023}. 
DeepTracer's proficiency extends to the accurate prediction of essential protein structure components, including amino acid positions, backbone location, secondary structures, and amino acid types \cite{Chang_Wang_Connolly_Meng_Su_Cvirkaite-Krupovic_Krupovic_Egelman_Si_2022, Pfab_Phan_Si_2020}. However, despite its robust capabilities, DeepTracer exhibits limitations in handling low-resolution maps \cite{Giri_Roy_Cheng_2023}. The model approaches protein structure prediction through a combination of segmentation and classification, relying heavily on the presence of distinct features within the cryo-EM map for accuracy. In scenarios where these features are blurred or indistinct, DeepTracer's effectiveness diminishes. In our effort to develop DeepTracer-LowResEnhance, we focus on pre-processing enhancements to cryo-EM maps, aiming to bolster DeepTracer's performance by improving feature definition within the input maps.

\textbf{ModelAngelo}: The recent development of ModelAngelo, a GNN-based machine-learning approach for automated atomic model building in cryo-EM maps, further exemplifies the practical utility of GNNs. ModelAngelo enhances the accuracy of atomic models for proteins and nucleotides, marking a significant achievement in cryo-EM structure determination. Utilizing GNNs and an attention mechanism similar to AlphaFold, ModelAngelo integrates density information, sequence data, and neighboring nodes to address common issues in de novo methods and sequence prediction \cite{ModelAngelo}. A notable feature of ModelAngelo is its ability to fine-tune backbone geometry using torsion angles, which has been shown to yield superior backbone root-mean-square deviation (RMSD) compared to the likes of DeepTracer \cite{Si2023}. This improvement is due to the complex and non-linear interactions of amino acids in proteins, driven by strong pairwise potentials that define the forces between amino acid pairs. Research by Park and Yoon on Graph Neural Networks provides two key insights: first, the average unique node degree is crucial for GNNs to generalize to new graphs; second, GNNs offer better inference accuracy compared to traditional algorithms, especially in scenarios with significant pairwise potentials. \cite{Park_Yoon_2021}. However, our observations reveal a key limitation of ModelAngelo: its performance and accuracy in residue-matching are significantly contingent on the quality of the cryo-EM map \cite{Si2023}. In preliminary tests with low-resolution maps (ranging from 4 to 7 Å), ModelAngelo often struggled to generate accurate predictions, whereas DeepTracer displayed a higher success rate. This suggests that when the node-edge relationships in a cryo-EM map are not clearly defined, the performance of GNN-based models like ModelAngelo is adversely affected.

\subsubsection{Cryo-EM Map Enhancement}
Since map-to-model methods predict the protein structures according to their cyro-EM maps, it usually requires pre-processing such as map enhancement before actually feeding the maps to prediction models. Recent years have witnessed substantial progress in the development of map enhancement tools aimed at refining the clarity and interpretability of cryo-EM maps.

\textbf{B-factor Correction:} B-factor correction \cite{b-factor}, also known as temperature factor or Debye-Waller factor correction, is essential for refining cryo-EM maps to enhance their interpretability and accuracy. By adjusting B-factors, this method reduces noise and emphasizes genuine signals, improving the visibility of structural features. This homogenization of the map’s density aids in clearer and more accurate model building and refinement. However, the primary limitation of the B-factor method is that it is applied globally across the entire map. Given that map quality is often heterogeneous, this uniform approach can result in non-optimal corrections, leading to over-sharpened and under-sharpened areas. The method's effectiveness depends on initial model quality, and there is also a risk of overfitting and lack of standardization across software packages, leading to inconsistent results and complicating comparisons.

\textbf{Local Deblur Method:} The local deblur method \cite{local-deblur} enhances cryo-EM maps by focusing on specific regions, applying targeted corrections to improve clarity and resolution. Unlike global techniques, it adjusts based on local density variations, enhancing fine structural details without significant noise. However, this method is computationally intensive and relies on the initial map quality. Poorly resolved areas or noisy maps may still lead to artifacts. Additionally, variability in algorithm performance and lack of standardization can result in inconsistent outcomes, complicating comparative studies and reproducibility.

\textbf{phenix.auto\_sharpen:} The \textit{phenix.auto\_sharpen} tool \cite{Terwilliger:ic5102} is a non-deep-learning algorithm developed for the automatic enhancement of cryo-EM maps through a process of sharpening. Distinguished by its novel approach, \textit{phenix.auto\_sharpen} enhances map quality by carefully balancing the observed detail and connectivity, independently of any model-based or interpretative analysis of the map. This optimization involves quantifying the map's details through the surface area of iso-contour surfaces and evaluating map-model correlation with models adjusted to zero B factors. Such a methodology validates the effectiveness of adjusted surface area optimization as a significant mechanism for map sharpening, thereby positioning \textit{phenix.auto\_sharpen} as a vital benchmark in our comparative analysis.

\textbf{DeepEMhancer:} Employing a 3D U-Net architecture, DeepEMhancer stands at the forefront of deep learning applications in cryo-EM map enhancement. It is uniquely trained on pairs of experimental maps alongside their model-sharpened counterparts, enabling it to post-process experimental maps by simultaneously performing masking and sharpening operations \cite{DeepEMhancer}. DeepEMhancer's ability to improve map resolution through deep learning showcases a significant leap forward in the field.

\textbf{CryoFEM:} Introduced in November 2023, CryoFEM represents the latest innovation in cryo-EM map enhancement, integrating sophisticated deep-learning algorithms with predictions from AlphaFold to outperform established tools like \textit{phenix.auto\_sharpen} and DeepEMhancer. Unlike traditional methods, CryoFEM leverages convolutional neural networks (CNNs) trained on high-resolution maps and can be applied to predict and enhance lower-resolution maps \cite{cryoFEM}. This approach not only sharpens the maps but also reduces noise and improves the signal-to-noise ratio, making it easier for downstream map-to-model tools to identify and model the underlying structures. CryoFEM's success demonstrates the transformative potential of combining advanced deep learning techniques for image enhancement with AlphaFold predictions in automating model building and fitting for cryo-EM maps' atomistic interpretation. 

\subsection{Sequence-assisted Methods for Protein Structure Prediction}

\subsubsection{Sequence-to-model Method}
\textbf{AlphaFold}: In 2021, Jumper et al. introduced AlphaFold, which makes it possible to predict protein structures by incorporating novel neural network architectures on sequence data. AlphaFold's new architecture embeds multiple sequence alignments and pairwise features, enabling accurate end-to-end structure prediction \cite{Jumper_Evans_2021}. Furthering this understanding, Bouatta et al.'s study emphasizes the core features of AlphaFold. Its architecture features a neural network 'trunk' and a structure module, which leverages attention mechanisms to manage long-range dependencies in protein sequences and structures. Initially developed for natural language processing, the attention mechanism in AlphaFold plays a pivotal role in capturing complex interactions within proteins \cite{Bouatta_Sorger_AlQuraishi_2021}. The study accentuates the importance of attention mechanisms in managing complex data like protein sequences and structures. 

\subsubsection{Combined Methods}
In 2022, Terashi et al. explored the application of AlphaFold for refining protein models derived from cryo-EM maps. They introduced the DAQ-refine protocol, which uses AlphaFold for local refinement of protein models, particularly in low-quality regions \cite{Terashi_Wang_Kihara_2022}. This study not only showcased the utility of AlphaFold in enhancing structural accuracy but also highlighted the role of deep learning in evaluating and refining protein models. DeepTracer-Refine, a recent research at DAIS Group, also exhibits the advancement of integrating AlphaFold. DeepTracer-Refine automates the alignment of AlphaFold's structure with DeepTracer’s model, improving residue coverage and overall accuracy \cite{Chen_Zia_Wang_Hou_Cao_Si_2023}. DeepTracer-Refine illustrates the synergy between refinement methods and automated machine learning approaches, furthering the potential of attention techniques in protein structure prediction.

\subsection{Limitations of Current Methods}
Although current methods are adept at handling high-resolution cryo-EM maps, none of the above methods can effectively predict protein structure with low-resolution cryo-EM maps. Although AlphaFold only requires sequence data, it struggles with fold-switching regions in proteins \cite{Chakravarty_Porter_2022}. With the increasing number of low-resolution maps available, the development of automated tools to combine the potential of low-resolution maps and AlphaFold is an urgent and vital task in advancing the field of protein structure prediction.

%% file: tex/method.tex
\section{Methodology}
 
This chapter delineates our approach to advancing the field of protein structure prediction by blending established methods with engineering solutions. We aim to integrate the robust attributes of existing systems while infusing novel techniques to enhance prediction accuracy and efficiency. Our proposed methodology introduces AlphaFold at an earlier stage to enhance the features in low-resolution maps with its sequence data. This approach stands in contrast to traditional map sharpening tools such as \textit{phenix.auto\_sharpen}, which rely on overall B factors and evaluate map interpretability through surface area and connectivity of iso-contour surfaces \cite{Afonine_Klaholz_Moriarty_Poon_Sobolev_Terwilliger_Adams_Urzhumtsev_2018, Terwilliger:ic5102}. Our strategy employs deep learning techniques to fit a simulated map from AlphaFold with its corresponding cryo-EM map, subsequently utilizing this refined map to guide DeepTracer prediction. 

\begin{figure}[h]
\centering
\vspace{10pt} 
\includegraphics[width=0.9\linewidth]{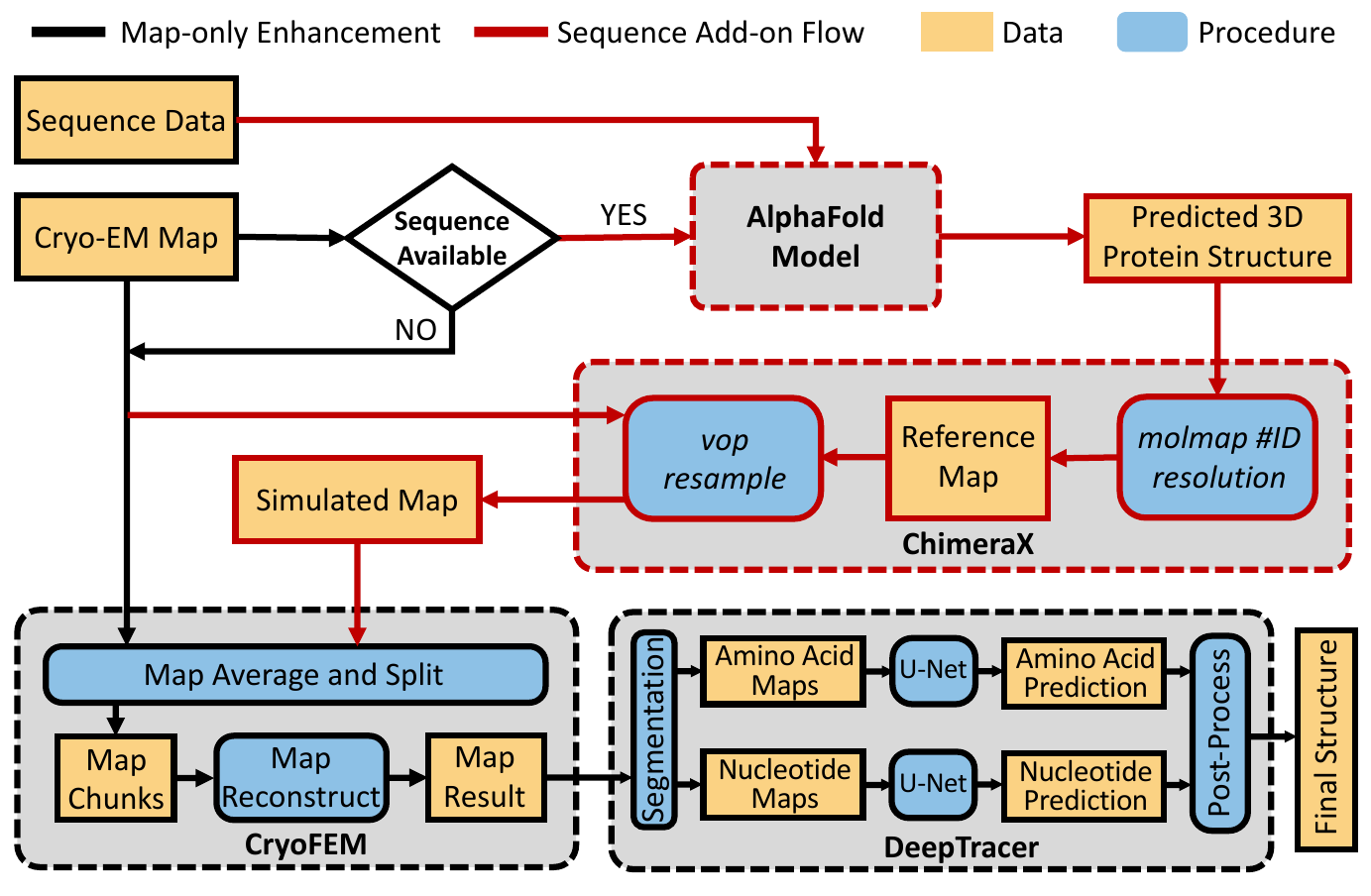} 
\caption{Workflow of DeepTracer-LowResEnhance. The process begins with the input of a cryo-EM map, optionally accompanied by sequence data. If sequence data is available, it is processed through AlphaFold for an initial 3D structure, refined using ChimeraX, which generates a simulated map. Both maps are then fed into the CryoFEM module, where the simulated map and the cryo-EM map are averaged and split into map chunks. A map reconstructing deep neural network will generate the refined map, which is then processed by DeepTracer to generate a high-accuracy 3D protein structure model.}
\label{fig:workflow}
\vspace{10pt} 
\end{figure}

\subsection{Workflow} 

Fig. \ref{fig:workflow} illustrates our modeling workflow, which represents an engineered approach to protein structure prediction, leveraging the strengths of existing computational models while introducing innovative methodologies for enhanced accuracy and efficiency. The process begins with the mandatory input of a cryo-EM map and optionally includes its corresponding sequence data. A sequence availability check determines the subsequent steps. If sequence data is available, it is processed through the AlphaFold model to predict an initial 3D protein structure. ChimeraX\cite{Pettersen_Goddard_Huang_Meng_Couch_Croll_Morris_Ferrin_2020} is used to generate a reference map from the AlphaFold model, match voxel dimensions of the cryo-EM map, and then export to a simulated map. This simulated map and the cryo-EM map are then processed through the CryoFEM module, which calculates the average of two maps, splits the result into chunks, reconstructs using a deep learning model, and produces a refined result using the \textit{phenix.dock\_and\_rebuild} tool \cite{cryoFEM}. In scenarios lacking an AlphaFold model, the CryoFEM map enhancement process commences directly with the raw cryo-EM map. The refined map is further processed by DeepTracer, which segments it into amino acid and nucleotide maps and then undergoes specialized U-Net analysis for prediction. The final post-processing phase involves refining predicted nucleotide structures and identifying double-stranded helical motifs. The combination of these steps results in the generation of a final, refined 3D protein structure model.

\subsection{Data Collection and Experiments}

To validate the effectiveness of our proposed method, we carefully select 37 test data from the Electron Microscopy Data Bank (EMDB) \cite{lawson2016emdatabase} with a specific criteria, which focuses on 22 entries with a reported resolution range of 4.0 to 8.4 Å. We also include 15 higher-resolution maps (2.5 - 4.0 Å) to cover a wider resolution range. The experiment process involves the following steps:

\begin{enumerate}
    \item Baseline model generation by directly processing cryo-EM maps through DeepTracer.
    \item Generation of DeepTracer predicted models from cryo-EM maps enhanced by \textit{phenix.auto\_sharpen}.
    \item Acquisition of AlphaFold models.
    \item Generating predictive models through the DeepTracer-LowResEnhance framework, selectively employing AlphaFold simulated maps in optimal scenarios and reverting to the original cryo-EM maps in cases identified as constraints.
    \item Compare our predicted model structure with the ground truth model structure from the PDB database. Evaluation of predictions employing Phenix tools \cite{Liebschner2019} to assess overall map-model correlation, total residues, and residues within acceptable density parameters.
\end{enumerate}

\subsection{Evaluation Metrics}

The performance and effects of a predicted protein model rely on several critical metrics that collectively estimate its accuracy and comprehensiveness. Therefore, three primary metrics are integrated into our study: map-model correlation, total residues, and residues in acceptable density.

\textbf{Overall Map-model Correlation:} For measuring the alignment between predicted models and the original cryo-EM maps, we utilize the map-model correlation coefficient (CC) tool provided by Phenix. CC is a statistical measurement that quantifies the level of correspondence between the electron density map and the molecular structure of the predicted model. This correlation coefficient ranges from -1 to 1, where:

\begin{itemize}
    \item A value of 1 means a perfect match, indicating that the model and the experimental data are in complete harmony.
    \item A value of 0 means no correlation, suggesting that the model does not reflect any meaningful relationship with the electron density map.
    \item A value of -1, although uncommon in structural biology, would imply a perfect inverse correlation, indicating that the model diametrically opposes the map.
\end{itemize}

Ideally, a high positive correlation coefficient close to 1 suggests that the atomic model is a good representation of the electron density map, implying that the positions of atoms and the shape of molecular structures in the model closely match those inferred from the experimental data. Conversely, a low correlation coefficient suggests discrepancies between the model and the map, indicating that the model may need refinement or that the cryo-EM map might be of insufficient resolution to accurately determine the protein features. The Phenix developers mentioned that while correlation coefficients can highlight regions of a model that may not align well with the map, these coefficients should not be viewed as definitive scores. Furthermore, identifying a universal threshold for these scores is challenging, as they do not consistently correlate with model resolution \cite{Liebschner2019}. Another investigation supports this, noting that CC scores do not always directly correlate with resolution and higher scores can occur in lower-resolution maps \cite{Pintilie_Chiu_2018}. Therefore, we treat the map-model correlation coefficient as a relative metric, using its proximity to 1 or 0 as an indicator of stronger or weaker correlation, respectively, rather than as a definitive measure of model quality.

\textbf{Total Residues and Residues in Acceptable Density:} In our proposed method, we employ the Phenix tool to rigorously assess both the total number of residues and their electron density quality within predicted models. The Phenix tool normalizes electron density values across the model's regions, setting the mean to zero and the standard deviation to one. This normalization process allows for consistent comparison and evaluation of electron density throughout different model regions, thereby standardizing the thresholds for density quality categories. The evaluation of residues in the predicted model is guided by the following criteria, as defined by Phenix \cite{Liebschner2019}:

\begin{itemize}
    \item \textbf{Out of Density:} Refers to atoms or groups positioned in areas where the electron density is significantly low, specifically more than two standard deviations below half the mean density for that atom or group. This indicates a probable absence of experimental evidence supporting the presence of these atoms or groups in those locations.
    \item \textbf{Very Weak Density:} Describes atoms or groups in regions where the electron density is one standard deviation below half the mean density. Though some density is present, it is insufficiently reliable, suggesting these model parts may be inaccurately placed or in unsupported conformations.
    \item \textbf{Weak Density:} This category applies to atoms or groups situated in areas where the electron density is below half the mean density, implying that, while there is evidence for the atoms' or groups' positions, the confidence in their precise placement is lower.
    \item \textbf{Acceptable Density:} Atoms or groups are supported by sufficient electron density, with no atom falling more than one standard deviation below half the mean density for that atom type, and no group exhibiting density less than half the mean for that group. This suggests a high level of confidence in these parts of the model.
\end{itemize}

In evaluating model quality and its fitness with the cryo-EM map, we place greater emphasis on the metric of Residues in Acceptable Density instead of traditional metrics like map-model correlation coefficients or total residue counts. This prioritization is based on the assumption that Residues in Acceptable Density offer a more direct reflection of the model's alignment with the electron density map. While map-model correlation coefficients offer a global measure of how well the model corresponds to the overall density map, they do not provide specific insights into the local accuracy of atomic positions. Similarly, the total number of residues does not account for the distribution of those residues across different density quality categories. Thus, we use map-model correlation and total residues as references. 

Upon establishing the count of Residues in Acceptable Density, we proceed to compute the percentage increase in this metric between the model under evaluation and the established DeepTracer baseline tool, as shown in Equation~\ref{residue_percentage_increase}. This percentage increase provides a quantitative evaluation of the comparative model's performance relative to the baseline, thereby enabling a more objective measurement of the predicted model's quality. This analytical strategy ensures that our evaluations are based on the essential elements of structural accuracy and empirical substantiation, fostering a solid and engineering-focused approach to model quality assessment.

\begin{equation}
    \begin{split}
    \text{Percentage Increase} = \frac{\text{Comparison Model} - \text{Baseline Model}}{\text{Baseline Model}} \times 100\%
    \end{split}
    \label{residue_percentage_increase}
\end{equation}

\textbf{Similarity Comparison with Solved Structure:} To validate the predictions made by DeepTracer-LowResEnhance, we align our results with solved structures from the Protein Data Bank (PDB) to get the TM-score using MultiMer-align (MM-align) \cite{MM-align}. MM-align is specifically designed for protein complexes alignment, which compares predicted protein complexes against their solved counterparts. Unlike traditional protein-structure alignment tools that might struggle with the unique challenges when handling multimeric proteins, MM-align excels by maintaining the integrity of protein-protein interfaces and accurately reflecting global structural similarities.

The TM-score, utilized by MM-align as a metric of structural superposition accuracy and coverage, is particularly effective for comparing protein structure similarity. It ranges from 0 (no similarity) to 1 (identical structures), where higher scores indicate a closer match between the predicted and solved structures. Specifically, a TM-score above 0.5 suggests that the structures share the same fold, which means a high degree of similarity. Scores less than 0.17 indicate close to random structural similarity \cite{MM-align}. This normalization feature of TM-score, accounting for protein length, offers a more consistent and reliable metric of structural similarity than conventional RMSD metric, especially when dealing with protein complexes of varying chain lengths.

We chose MM-align over alternatives like ChimeraX MatchMaker \cite{Pettersen_Goddard_Huang_Meng_Couch_Croll_Morris_Ferrin_2020} because it specializes in aligning quaternary structures and can handle complex multichain alignments without creating unrealistic cross-chain connections. By leveraging MM-align's strengths, we can more accurately evaluate the performance of DeepTracer-LowResEnhance against solved PDB structures. This comparative analysis not only validates the accuracy of our predicted models but also reinforces our confidence in applying DeepTracer-LowResEnhance to newly acquired low-resolution maps, thereby advancing the field of structural biology.

\subsection{Assumptions and Limitations}

\begin{figure}[!h]
  \centering
  \includegraphics[width=0.6\linewidth]{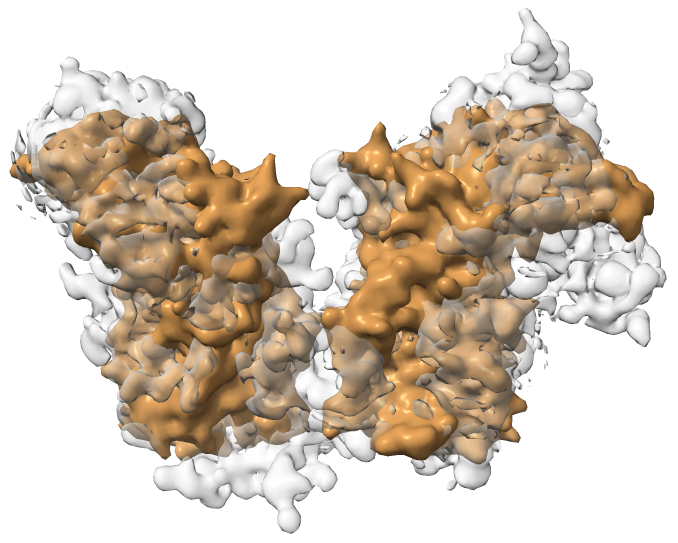}
  \caption{A depiction of a limiting scenario with EMD-29072: A 6.0 Å resolution cryo-EM map in gray juxtaposed against its AlphaFold simulated counterpart in brown, highlighting the critical potential for shape mismatch errors.}
  \label{fig:edge_case_29072}
\end{figure}

To improve protein structure prediction from cryo-EM maps, our proposed method leverages the integration of simulated maps derived from AlphaFold models. This section highlights when integrating AlphaFold is most beneficial and identifies situations where it might face challenges or be less effective. By identifying these contexts, we provide a clear guide to help researchers use our methods more precisely and effectively.

\textbf{Optimal Scenarios:} DeepTracer-LowResEnhance performs best with cryo-EM maps in the 4.0 to 8.5 Å resolution range. It typically processes a cryo-EM map, optionally combined with a simulated map from an AlphaFold model, to improve the clarity and accuracy of low-resolution regions. The AlphaFold model should cover the region of interest in the cryo-EM map, and the simulated map should be created according to the guidelines in section 3.1.

\textbf{Constraint Scenarios:} However, there are situations where adding a simulated map may not help and lower its performance. These situations usually occur when the original cryo-EM map has a high resolution (better than 4.0 Å), making the simulated map unnecessary. Additionally, if there are mismatches between the AlphaFold model's predicted conformation and the actual protein conformation in the cryo-EM map, it can cause inaccuracies. For example, Fig. \ref{fig:edge_case_29072} shows a discrepancy between a 6.0 Å resolution cryo-EM map (in gray) and its simulated analog (in brown) for EMD-29072. In such cases, relying only on the original cryo-EM map is better. Moreover, for extremely low-resolution maps (below 9.0 Å), the enhancement process is less effective. Although some resolution improvement is possible, accurately predicting a significant number of residues in the map-to-model conversion pipeline is very challenging.

%% file: tex/results.tex
\section{Results}
This chapter presents the evaluation experiments of our proposed method, highlighting the significant advancements made in protein structure prediction using low-resolution cryo-EM maps.

\subsection{Improvements in Handling Low-resolution Maps}

\begin{figure}[h]
  \centering
  \includegraphics[width=0.7\linewidth]{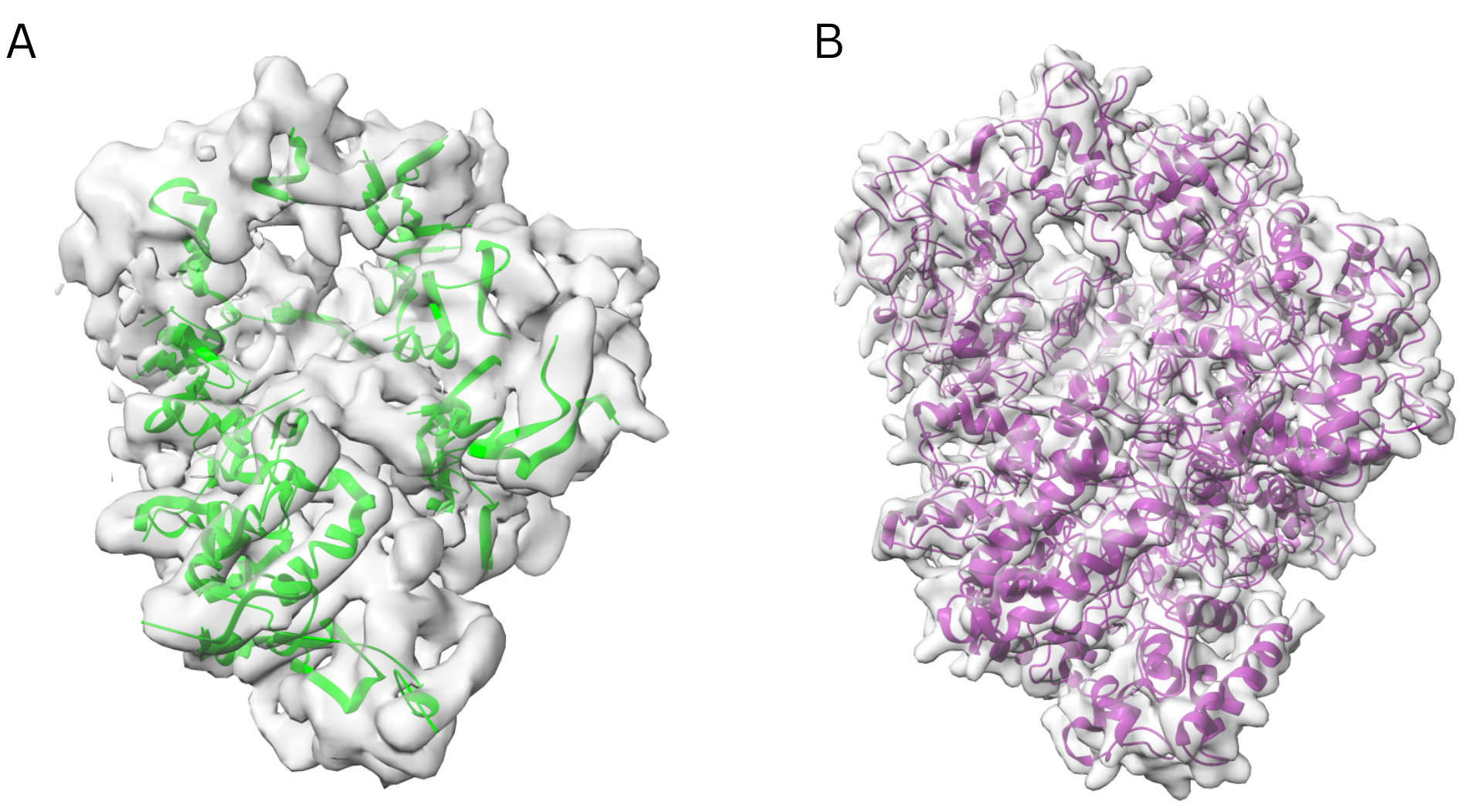}
  \caption{A comparison of DeepTracer's low-resolution map prediction results. \textbf{A:} Predictions from the 6.7 Å cryo-EM map (EMD-4141) using DeepTracer (green) showing only 369 residues. \textbf{B:} DeepTracer-LowResEnhance’s predictions (purple) display a significant increase to 1832 residues.}
  \label{fig:emd_4141_result}
\end{figure}

We demonstrate that integrating a map treatment process with DeepTracer-LowResEnhance significantly enhances its capability to process lower-resolution maps, especially those ranging from 4.0 to 8.4 Å. The comparison of total residues in atomic models predicted by the original DeepTracer and those refined by DeepTracer-LowResEnhance across 22 low-resolution maps shows notable improvements in 95.5\% of cases, with only one map not showing enhancements. 
The increase in total predicted residues is notably illustrated in the analysis of the 6.7 Å resolution map EMD-4141, PDB-ID: 5M1S (Fig. \ref{fig:emd_4141_result}). The original DeepTracer tool predicates only 369 residues from the unprocessed map in 38 seconds, with 277 of those residues meeting the criteria for acceptable density (Fig. \ref{fig:emd_4141_result}A). In contrast, using DeepTracer-LowResEnhance with sequence data leads to a substantial increase in the number of predicted residues to 1832, with 1187 of these residues within the acceptable density range (Fig. \ref{fig:emd_4141_result}B). This clearly showcases the significant impact of our map enhancement process, except for a slightly increased processing time of 3 minutes and 28 seconds. Additionally, the map-model correlation coefficient improved from 0.270 to 0.442, indicating a more robust alignment between the cryo-EM map and the predicted model.

\begin{figure}[!h]
\centering
\includegraphics[width=0.75\linewidth]{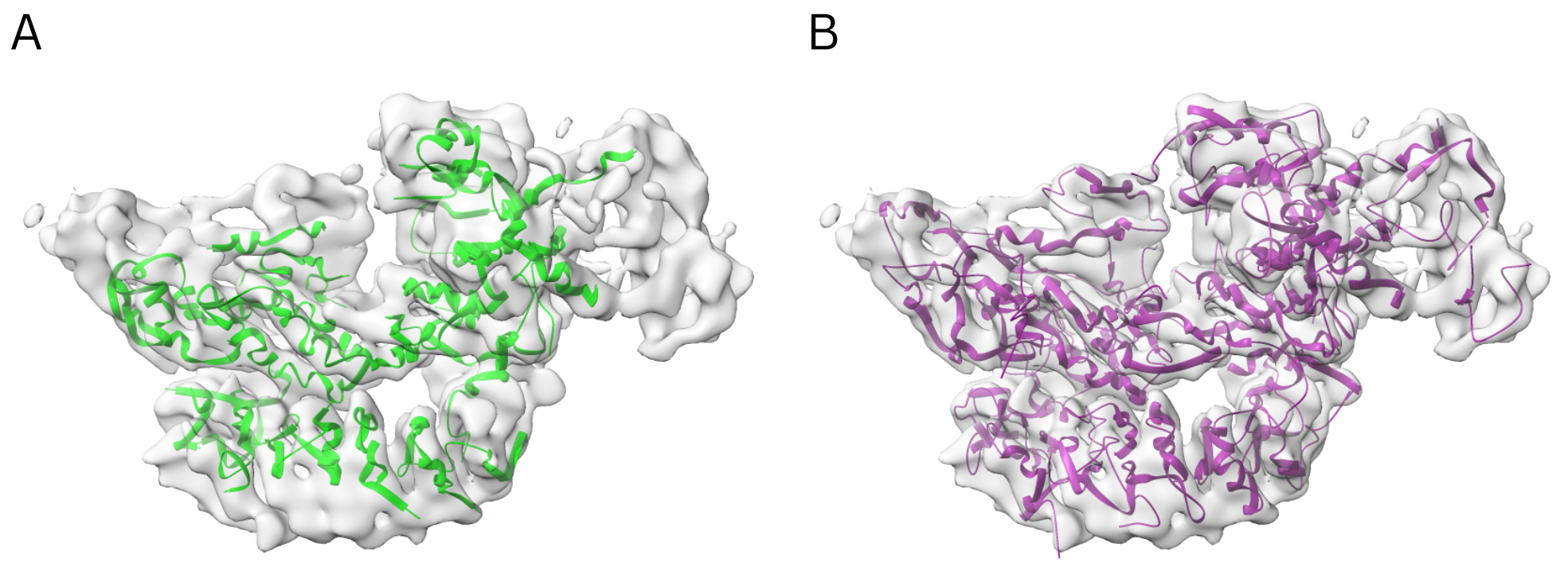}
\caption{Evaluating the efficacy of DeepTracer-LowResEnhance without integrating sequence data, using the 8.3 Å cryo-EM map, EMD-17124. \textbf{A:} Initial DeepTracer output shows 316 residues within acceptable density, 440 total residues, and a 0.175 map-model correlation. \textbf{B:} With DeepTracer-LowResEnhance, the figures rise to 592 residues within acceptable density, 817 total residues, and a 0.214 map-model correlation.}
\label{fig:emd_17124_result}
\end{figure}

Initially, our approach aimed to combine CNNs and GNNs into a unified modeling framework. However, through experimental validation, we found that while the GNN-based ModelAngelo struggled with maps of resolution poorer than 4 Å, the CNN-based DeepTracer performed much better in interpreting lower-resolution maps due to its architectural advantages. For example, DeepTracer successfully predicted 1390 residues in a 7.2 Å map (EMD-2183), whereas ModelAngelo could not produce a viable prediction. These critical insights lead us to choose DeepTracer as the preferred tool for processing low-resolution maps in our pipeline.

In specific instances, integrating AlphaFold simulated maps presents challenges such as shape mismatch errors (discussed in section 3.4) observed in 2 out of 22 low-resolution maps. Additionally, 3 maps lack the sequence data necessary for AlphaFold predictions. All 15 high-resolution maps underwent processing with AlphaFold simulated maps. As a result, 5 maps were processed exclusively with the original cryo-EM map. For instance, EMD-17124 in Fig.\ref{fig:emd_17124_result}, an 8.3 Å map lacking sequence data or a pre-existing PDB model, shows significant improvements in residues within acceptable density, total residue count, and overall map-model correlation upon treatment with DeepTracer-LowResEnhance. This underscores the tool's usefulness in refining low-resolution maps without AlphaFold predictions, enabling the use of new cryo-EM data that may lack sequence information or computational models. Consequently, this approach serves to narrow the existing gap between the volume of accessible maps and the quantity of solved protein structure models, as previously discussed in Section 1.

\begin{figure}[!h]
\centering
\includegraphics[width=0.8\linewidth]{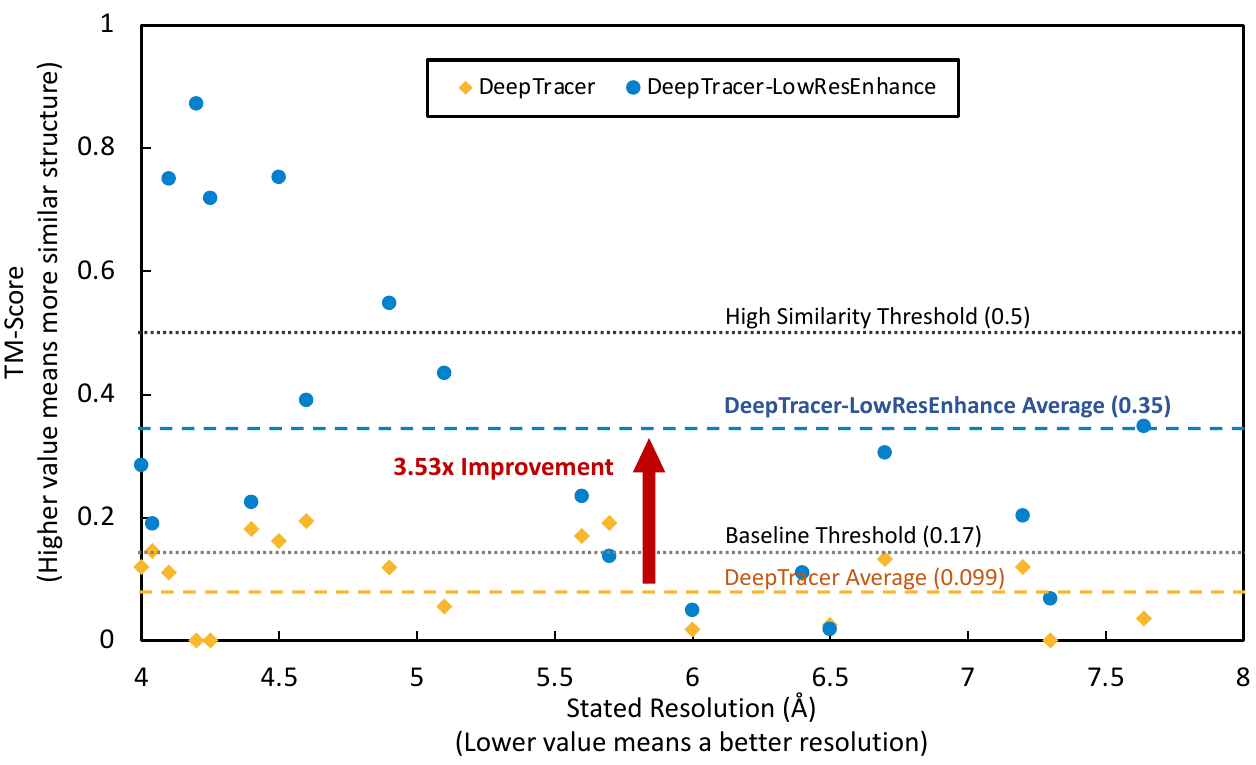}
\caption{Comparison of structure similarity between DeepTracer-LowResEnhance predictions and solved structures for low-resolution maps. The metric (TM-score) ranges from 0 to 1, with higher scores denoting greater structural similarity. Out of 19 tests, 14 surpass the baseline threshold of 0.17, indicating significant similarity beyond random coincidence. The DeepTracer tool averages a TM-score of 0.099 (below threshold) while our proposed method achieves an average TM-score of 0.35, marking a 3.53 times improvement over DeepTracer.}
\label{fig:structure_similarity_analysis}
\end{figure}

In tests comparing DeepTracer-LowResEnhance predictions with solved PDB structures using MM-align \cite{MM-align}, we evaluate 22 low-resolution maps, 19 of which have corresponding solved PDB structures. By deriving TM-scores through MM-align, we quantify the similarity between our predictions and the solved structures. The results reveal that the majority of our predictions from low-resolution maps show acceptable to high similarity with existing PDB structures, confirming our tool's reliability and its effectiveness for analyzing low-resolution cryo-EM data (Fig. \ref{fig:structure_similarity_analysis}). Instances where TM-scores fall below 0.5 often corresponded with a low count of predicted and matched residues, but still showing great improvement. For example, EMD-26563 initially had only 6 residues predicted by DeepTracer, this count increases to 58 with DeepTracer-LowResEnhance. However, it is still significantly less than the corresponding solved PDB structure. This discrepancy highlights the challenge of achieving high TM-scores with fewer predicted residues. 

\begin{figure}[!h]
  \centering
  \includegraphics[width=0.8\linewidth]{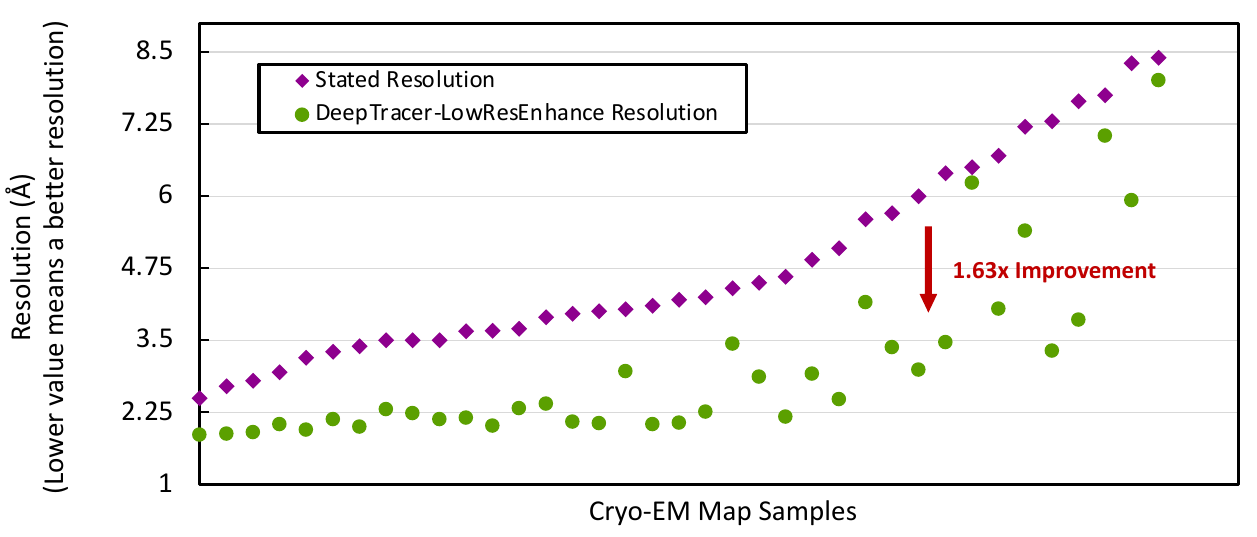}
  \caption{Comparison of stated and DeepTracer-LowResEnhance improved cryo-EM map resolutions, where lower values represent better map resolution. \textbf{Green points:} DeepTracer-LowResEnhance improved map resolution.  \textbf{Purple points:} stated map resolution. The analysis demonstrates consistent improvement across all evaluations, with an average enhancement of 1.63 times.}
  \label{fig:resolution_comparison}
\end{figure}

Furthermore, our comprehensive assessment across 37 datasets, utilizing Phenix Mtriage \cite{Afonine_Klaholz_Moriarty_Poon_Sobolev_Terwilliger_Adams_Urzhumtsev_2018}, highlighted universal improvements in map resolution. Fig. \ref{fig:resolution_comparison} shows the improved map resolution with DeepTracer-LowResEnhance in comparison with the stated resolution of the original cryo-EM map. The improvement in resolution is a good indicator of a higher level of detail visible in the cryo-EM map. It's important to acknowledge that not every enhanced cryo-EM map with a higher resolution resulted in better-predicted atomic models, particularly with maps already at high resolution. Challenges such as bias, the introduction of errors, and loss of structural context can arise, often attributed to over-sharpening. These issues highlight the need for an engineered approach in map enhancement processes and the importance of careful selection of approaches. We will explore these complexities and their implications further in Chapter 5 on these limitations. This advancement paves the way for more accurate and feasible structural analyses of low-resolution cryo-EM maps, thereby extending the reach of cryo-EM in structural biology.

\subsection{Comparison with \textit{phenix.auto\_sharpen}}

\begin{figure}[!h]
    \centering
    \begin{subfigure}[t]{0.8\textwidth}
        \centering
        \includegraphics[width=\linewidth]{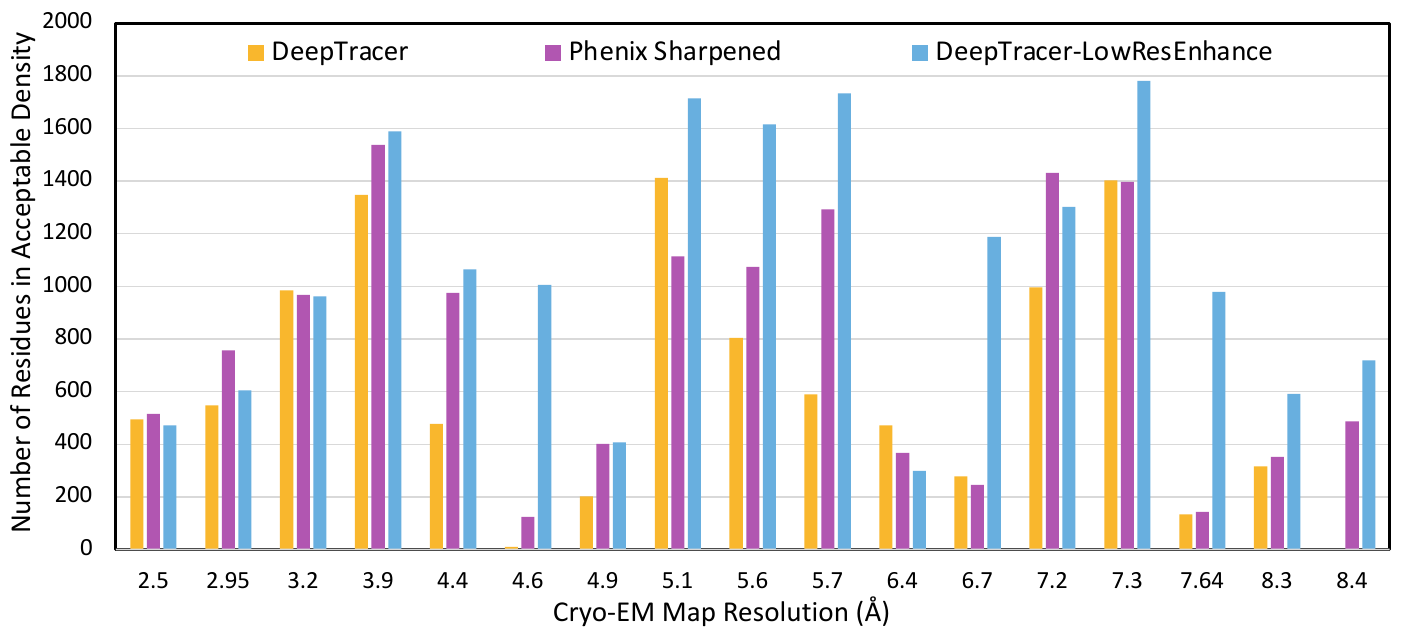}
        \caption{Comparative analysis of residues in acceptable density across different cryo-EM map resolutions. The performance of the DeepTracer baseline (in orange), phenix.auto\_sharpen (in purple), and DeepTracer-LowResEnhance (in blue) predictions are shown. The chart highlights DeepTracer-LowResEnhance identifies more residues within acceptable density than other methods, especially for cryo-EM maps with resolutions worse than 4 Å (higher value means worse resolution).}
    \label{fig:comparison_phenix_a}
    \end{subfigure}%

    \vspace{0.5cm} 
    
    \begin{subfigure}[t]{0.8\textwidth}
        \centering
        \includegraphics[width=\linewidth]{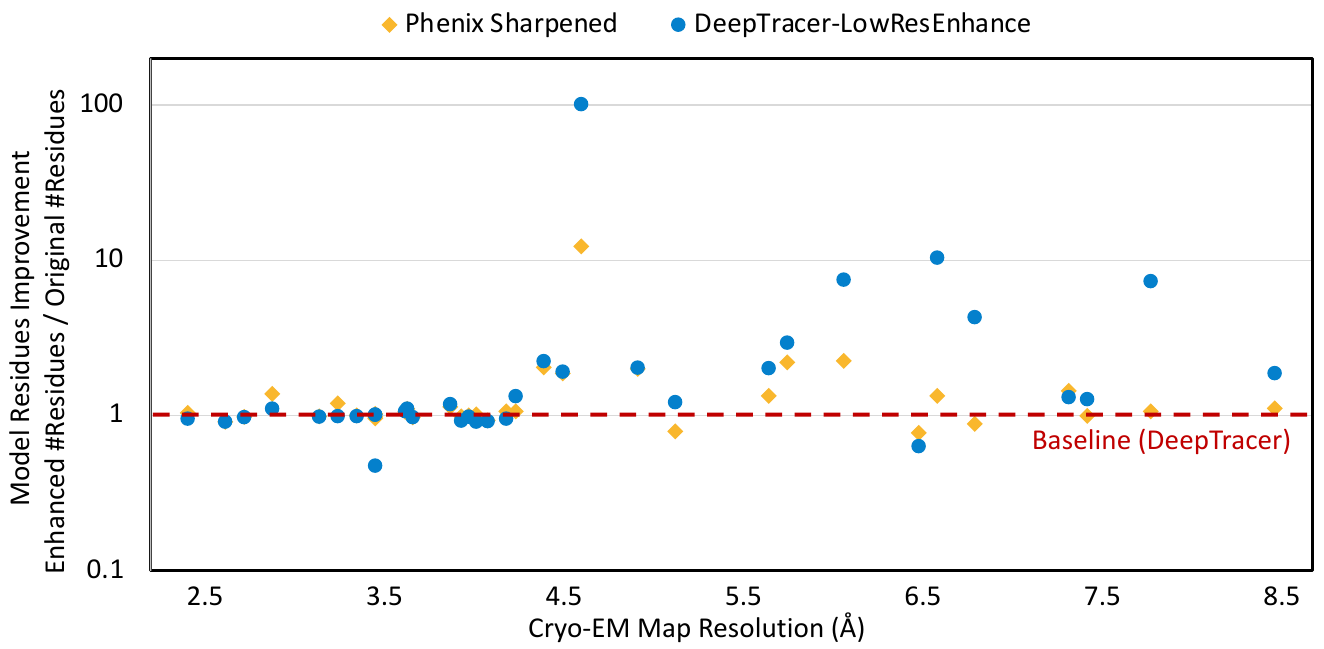}
    \caption{Residues in acceptable density of comparison tools with the DeepTracer baseline. The performance of phenix.auto\_sharpen (in orange) and DeepTracer-LowResEnhance (in blue) is compared against the baseline (red dashed line). Both phenix.auto\_sharpen and our proposed method show similar results to the baseline for higher-resolution (better than 4.5 Å) maps. However, our proposed method significantly outperforms the baseline and phenix.auto\_sharpen, particularly in lower-resolution (worse than 4.5 Å) cryo-EM maps.}
    \label{fig:comparison_phenix_b}
    \end{subfigure}
    \caption{A comparison of residues in acceptable density.}
    \label{fig:comparison_phenix}
\end{figure}

This section compares the performance of DeepTracer-LowResEnhance with \textit{phenix.auto\_sharpen} \cite{Terwilliger:ic5102} using a data-driven approach. Measuring improvements in residue identification within acceptable density ranges highlights how our tool enhances the analysis of low-resolution cryo-EM maps. This comparison provides clear evidence of the advanced capabilities and effectiveness of DeepTracer-LowResEnhance.

Fig. \ref{fig:comparison_phenix} presents a comparative analysis of residues in acceptable density across different resolutions, featuring the performance of the DeepTracer baseline (in orange), \textit{phenix.auto\_sharpen} (in purple), and DeepTracer-\\LowResEnhance (in blue) predictions. Our proposed method significantly outperforms \textit{phenix.auto\_sharpen} in identifying residues within acceptable density, especially in electron microscopy maps with resolutions worse than 4 Å (Fig. \ref{fig:comparison_phenix_a} and Fig. \ref{fig:comparison_phenix_b}). Specifically, for the 22 low-resolution maps analyzed, \textit{phenix.auto\_sharpen} achieves an average increase of 87.112\% in residues classified within acceptable density compared to the baseline, while DeepTracer-LowResEnhance achieves a remarkable 662.539\% improvement. However, this pronounced advantage of DeepTracer-LowResEnhance diminishes in the 15 analyzed high-resolution maps better than 4 Å (Fig. \ref{fig:comparison_phenix_b}), indicating its optimized performance is particularly beneficial for lower-resolution data. 

\subsection{Enhancing Cryo-EM Map Features through AlphaFold Integration}

\begin{figure}[!h]
\centering
\includegraphics[width=0.8\linewidth]{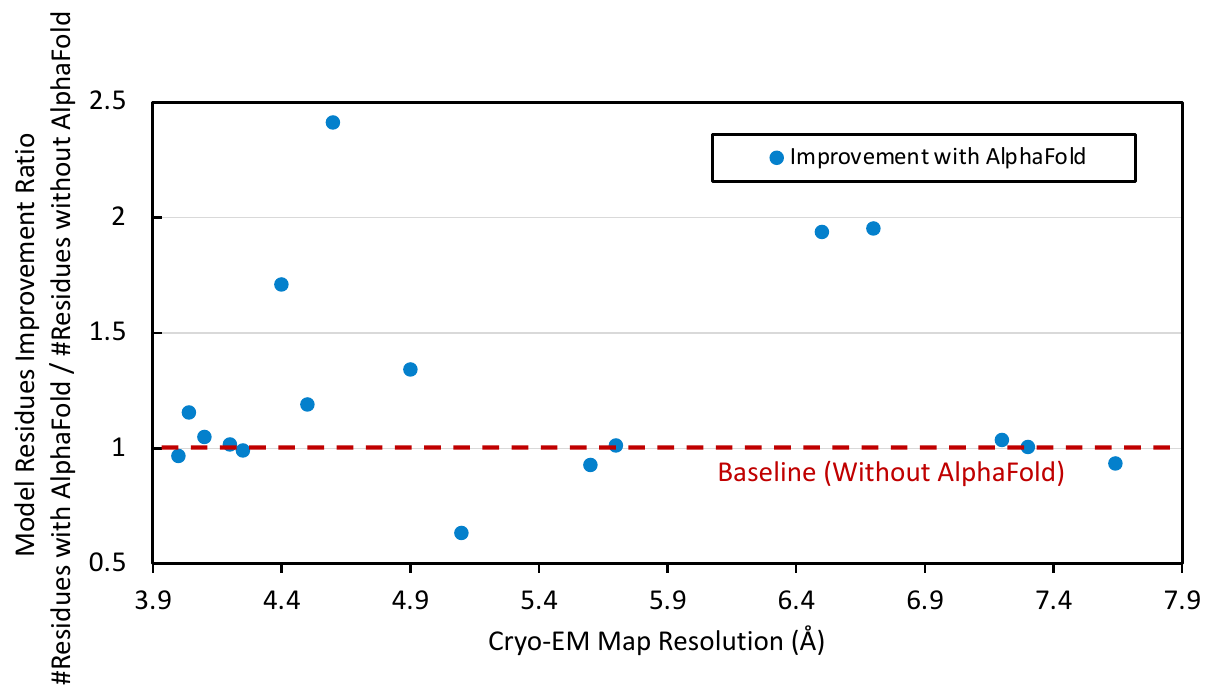}
\caption{Impact of AlphaFold integration on DeepTracer-LowResEnhance predictions. This figure showcases the comparative analysis of residue identification improvements within acceptable density by incorporating AlphaFold-generated simulations into the DeepTracer-LowResEnhance process. For the low-resolution maps analyzed, the combined approach (blue points) led to significant improvements compared to CryoFEM-enhanced maps feeding into DeepTracer without AlphaFold integration (red dashed line).}
\label{fig:af_integration}
\end{figure}

Integrating AlphaFold-generated simulations into our cryo-EM map enhancement process significantly improves upon direct map enhancement techniques in the DeepTracer-LowResEnhance workflow. By using AlphaFold simulations for the low-resolution maps analyzed, we enabled a detailed comparison with the outcomes from CryoFEM-processed maps before DeepTracer. As shown in Fig. \ref{fig:af_integration}, this combined approach led to notable improvements in identifying residues within acceptable density for most maps, with the exception of one deviation in map EMD-33798. These advancements were consistent across various metrics, including map-model correlation and total residue counts, highlighting the comprehensive benefits of incorporating AlphaFold in cryo-EM structure prediction. 

\begin{figure}[!b]
    \centering
    \includegraphics[width=0.3\linewidth]{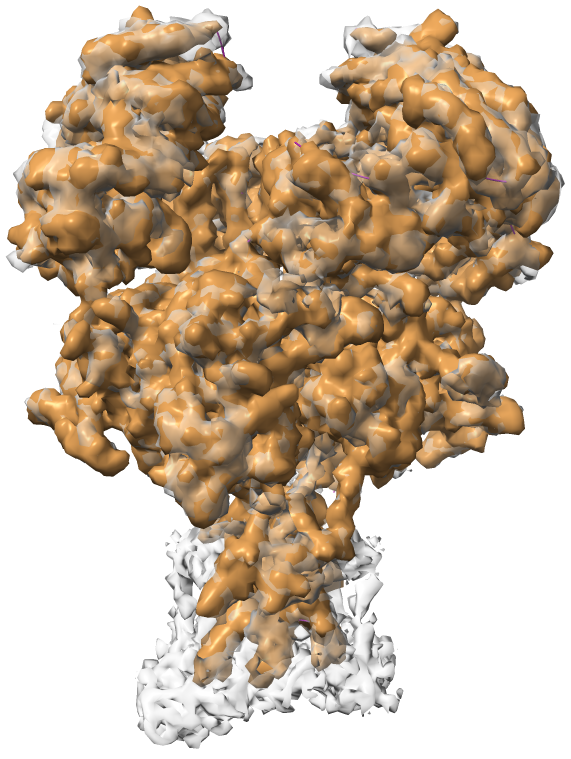}
    \caption{An illustration of EMD-33798: A 5.1 Å resolution cryo-EM map in gray juxtaposed against its AlphaFold simulated counterpart in brown reveals a significant discrepancy at the bottom part, emphasizing the challenges in accurately predicting the multiple conformational states of proteins. }
    \label{fig:analyze_emd_33798}
\end{figure}

Upon examining the outlier case of EMD-33798 (Fig. \ref{fig:analyze_emd_33798}), identified as the structure of the GluN1a E698C-GluN2D NMDA receptor in its cysteine non-crosslinked state, we delve into the complexity of protein conformations. Given that proteins may exhibit various conformational states reflective of their functional status, accurately predicting these states is paramount. However, AlphaFold's current capabilities do not extend to predicting multiple conformational states based solely on sequence data \cite{Skolnick_Gao_Zhou_Singh_2021}, which may account for discrepancies between AlphaFold predictions and cryo-EM map data. This observation highlights the inherent challenges in modeling proteins that exhibit multiple conformational states and underscores the need for continued advancements in predictive methodologies.

\subsection{Validation with Down-sampled High-resolution Cryo-EM Maps}

To validate the effectiveness of our proposed DeepTracer-LowResEnhance across varying cryo-EM map resolutions, we select two high-resolution cryo-EM maps: EMD-23019 (3.5 Å) and EMD-23807 (3.67 Å). These maps are manually down-sampled to different resolutions using ChimeraX. We then input the resulting low-resolution maps into both the original DeepTracer workflow and our proposed DeepTracer-LowResEnhance workflow to predict the corresponding PDB structures. To assess the accuracy of these predictions, we calculate the TM-scores by comparing the predicted protein structures against the reference structures from the PDB database.

\begin{figure}[!h]
    \centering
    \begin{subfigure}[t]{0.48\textwidth}
        \centering
        \includegraphics[width=\linewidth]{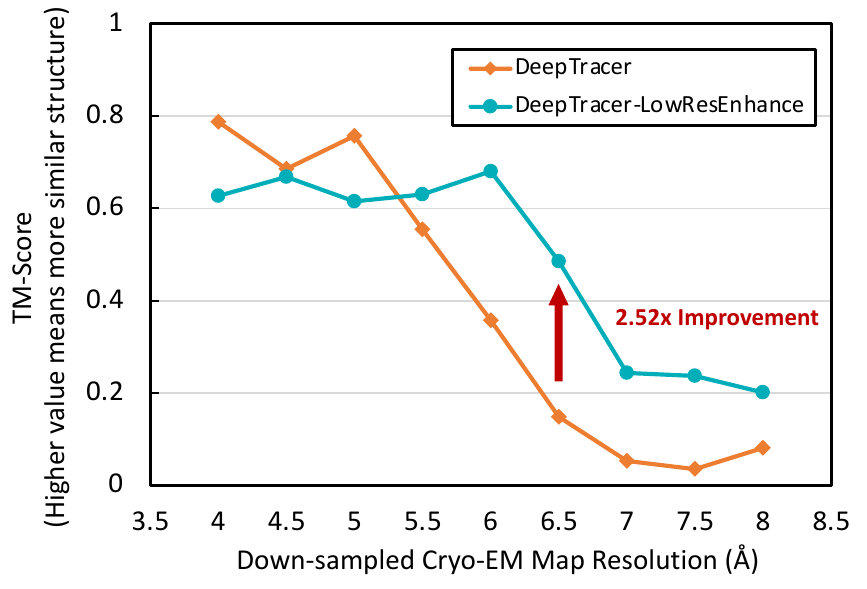}
        \caption{Validation results for a 3.5 Å resolution map EMD-23019 (PDB ID: 7KSL).}
    \end{subfigure}%
    ~ 
    \begin{subfigure}[t]{0.48\textwidth}
        \centering
        \includegraphics[width=\linewidth]{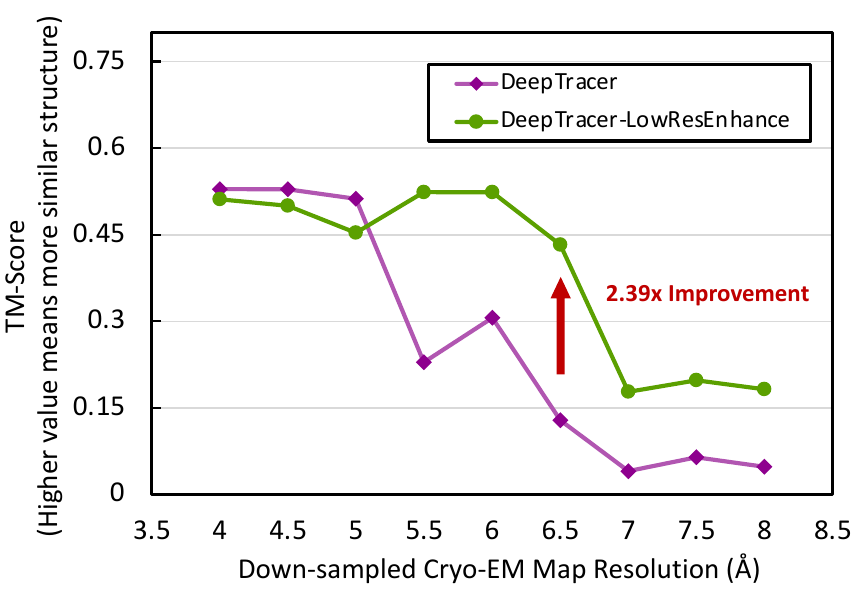}
    \caption{Validation results for a 3.67 Å resolution map EMD-23807 (PDB ID: 7MEY).}
    \end{subfigure}
    \caption{TM-score comparison of predicted protein structures from down-sampled cryo-EM maps at varying resolutions, showcasing the performance of DeepTracer-LowResEnhance relative to the original DeepTracer. Significant improvements are observed at resolutions worse than 5 Å.}
    \label{fig:validation_downsample}
\end{figure}

Fig. \ref{fig:validation_downsample} presents the TM-score results across the down-sampled maps at various resolutions. For maps with resolutions finer than 5 Å, both DeepTracer and DeepTracer-LowResEnhance achieve TM-scores above or close to the high-similarity threshold of 0.5. In this scenario, DeepTracer-LowResEnhance exhibits a slight reduction in TM-score, likely due to errors introduced from AlphaFold-generated simulations. However, as the resolution degrades beyond 5 Å, DeepTracer-LowResEnhance demonstrates substantial improvement over DeepTracer. On average, DeepTracer-LowResEnhance achieves a 2.52x improvement for EMD-23019 and a 2.39x improvement for EMD-23807 across all down-sampled resolutions.

\subsection{Comparison with CryoFEM}

\begin{figure}[!h]
    \centering
    \includegraphics[width=0.5\linewidth]{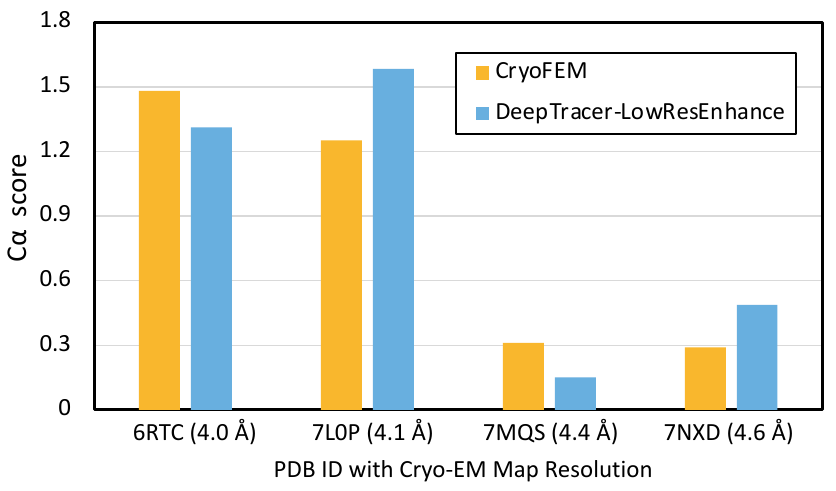}
    \caption{Comparison with CryoFEM workflow proposed in the CryoFEM paper\cite{cryoFEM} by calculating the $C_\alpha$ scores of the predicted protein structures following the methodology outlined in their paper. The comparison is based on four available data points in CryoFEM with the worst map resolution, where a higher $C_\alpha$ score reflects better model accuracy. DeepTracer-LowResEnhance demonstrates superior performance on two samples, with a slight reduction on the remaining two.}
    \label{fig:compare_CryoFEM}
\end{figure}

We also attempt to compare our proposed DeepTracer-LowResEnhance with the protein prediction workflow described in the CryoFEM paper\cite{cryoFEM}. However, the open-source code for CryoFEM only includes the map enhancement component, preventing us from reproducing their full process for generating predicted 3D protein structures. Additionally, the results in the CryoFEM paper are limited to high-resolution maps ranging from 2.8 Å to 4.6 Å, leaving us with insufficient low-resolution data for a direct comparison.

In our attempt, we select four relatively low-resolution maps (worse than 4 Å) from their dataset and calculate the $C_\alpha$ scores of the predicted protein structures following the methodology outlined in their paper. A higher $C_\alpha$ score indicates better model accuracy. The comparison results, shown in Figure \ref{fig:compare_CryoFEM}, indicate that DeepTracer-LowResEnhance delivers superior performance for the 7L0P (EMD-23099) and 7NXD (EMD-12637) protein samples while exhibiting a modest reduction in performance for the 6RTC (EMD-4997) and 7MQS (EMD-23951) proteins.

\subsection{Advancements Beyond AlphaFold's Capabilities}

\begin{figure}[!h]
\centering
\includegraphics[width=0.9\linewidth]{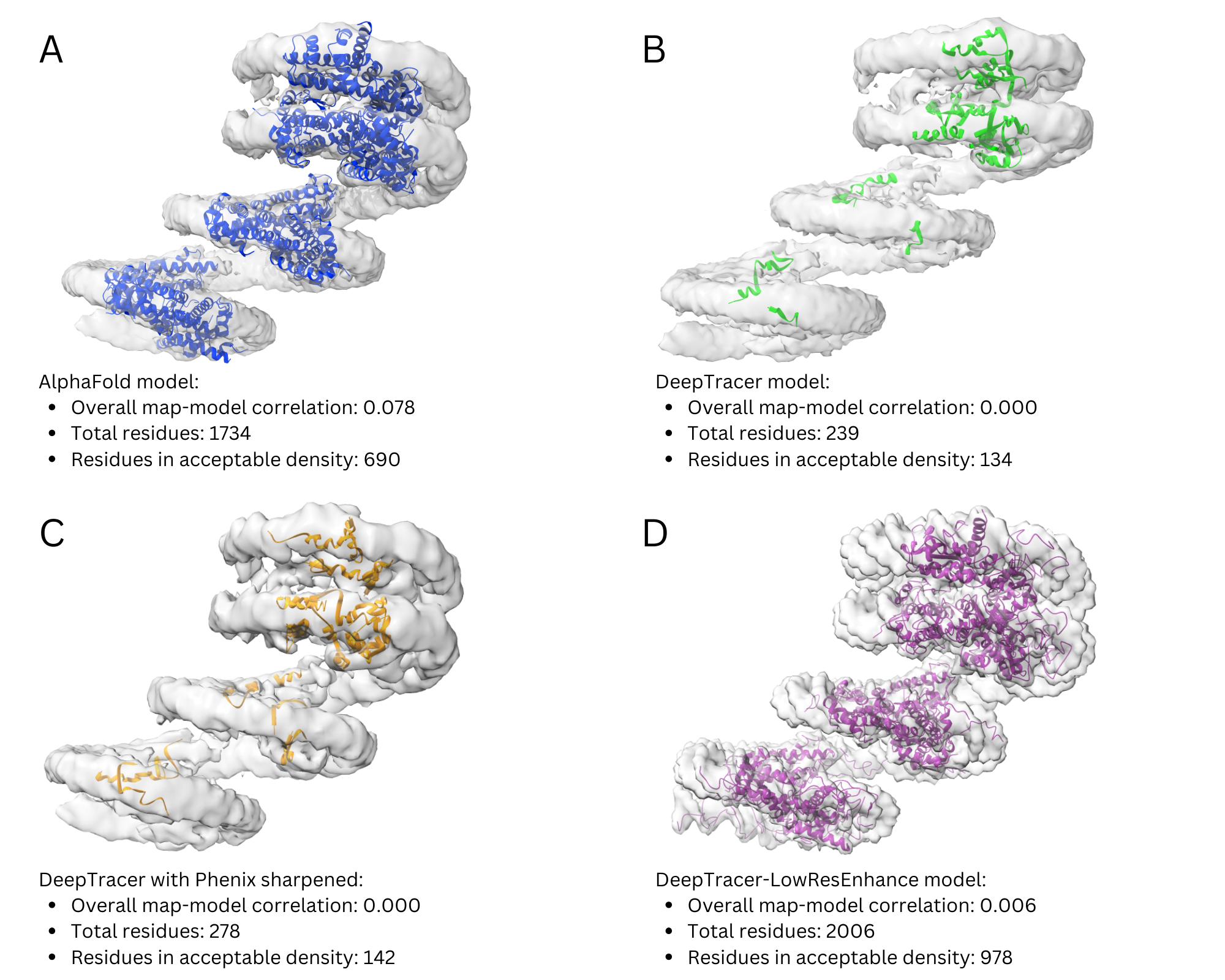}
\caption{Comparative analysis of prediction outcomes for EMD-35448 (7.64 Å), illustrates performance differentials among various tools. \textbf{A:} The AlphaFold tool, indicates a high map-model correlation yet a diminished count of total residues and those within acceptable density. \textbf{B:} Baseline prediction by DeepTracer. \textbf{C:} Enhancements rendered by \textit{phenix.auto\_sharpen}, showing negligible improvements over the baseline. \textbf{D:} DeepTracer-LowResEnhance's prediction, highlighting a pronounced elevation in the count of residues classified as within acceptable density.}
\label{fig:comparison_AF_emd_35448}
\end{figure}

In Section 2.3, we discussed the constraints of AlphaFold in precisely modeling regions of proteins that undergo fold switching. Such limitations often result in AlphaFold-generated models exhibiting fewer residues within acceptable density levels, reflecting a suboptimal alignment between the predicted model and the actual cryo-EM map data. Our approach shows a significant enhancement in overcoming these challenges. Fig. \ref{fig:comparison_AF_emd_35448} illustrates the advanced predictive performance of DeepTracer-LowResEnhance when applied to the 7.64 Å resolution cryo-EM map of EMD-35448. In this comparative analysis, DeepTracer-LowResEnhance not only outperforms the capabilities of both the DeepTracer and \textit{phenix.auto\_sharpen} but also evidences a notable advancement over the conventional AlphaFold tool. Specifically, DeepTracer-LowResEnhance identifies a substantially greater number of residues in acceptable density (978 compared to AlphaFold's 690), highlighting its superior proficiency in accurately describing critical structural features.

%% file: tex/future.tex
\section{Future Directions}

Our advancements in enhancing protein structure prediction, particularly through DeepTracer-LowResEnhance, have illuminated the path to addressing significant challenges within structural biology. While we've made notable strides in processing lower-resolution maps, our exploration has also uncovered areas requiring further development and innovation.

\textbf{High-Resolution Map Analysis:} Our interventions have shown limited effectiveness on maps finer than 4 Å, with some instances even seeing a decline in performance. This highlights a crucial need to enhance our methodologies for high-resolution challenges, particularly to mitigate over-sharpening effects. At these higher resolutions, the model accuracy is crucial, making any discrepancies between the map and model more apparent. Efforts to sharpen maps aim to improve feature visibility by enhancing high-frequency components, yet this might lead to over-sharpening and introduce artifacts that are not present in the original map. Such artifacts may be misinterpreted as genuine structural features by prediction algorithms, leading to model inaccuracies and biases. This observation emphasizes the need for an engineered approach to map sharpening that enhances detail without distorting the original data's integrity, highlighting the essential requirement to refine our approaches for high-resolution map analysis.

\textbf{DNA/RNA Structure Prediction from Low-Resolution Maps:} The unexplored territory of applying DeepTracer-LowResEnhance to DNA/RNA structures from low-resolution cryo-EM maps presents a promising avenue. Given that numerous model prediction tools have demonstrated proficiency in nucleotide prediction, assessing the effectiveness of DeepTracer-LowResEnhance in this area is a logical next step. Future investigations will assess its effectiveness in this domain, aiming to bridge a crucial gap in structural modeling.

\textbf{Enhanced Training with Larger Datasets:} The exponential growth of structural data repositories offers an opportunity to refine pre-trained models with expansive datasets, promising improvements in predictive accuracy and the capability to address complex or resolution-constrained structures.

\textbf{Graph Neural Networks Integration:} Exploring the integration of Graph Neural Networks (GNNs) could offer substantial enhancements in analyzing high-resolution maps, leveraging the unique advantages of GNNs in handling complex networked data like protein structures.

\textbf{Resolution Variation and Heterogeneous Distribution:} A pivotal area for future work involves addressing not just the overall resolution of cryo-EM maps but also the heterogeneous distribution of resolutions within a single map. Developing techniques to differentiate and enhance local and global resolutions will be crucial for accurate model prediction across varying map qualities.

\textbf{Optimized Usage of AlphaFold Based on Confidence Scores:} Future efforts should also focus on maximizing the use of AlphaFold by considering its confidence scores for local structures and map-model agreements. Instead of disregarding AlphaFold predictions that do not perfectly align with the map, we could have an engineered approach that leverages AlphaFold's predictions based on its reliability scores, ensuring a more integrated and informed analysis.

\textbf{Conformational Change Prediction:} Addressing the challenge of predicting proteins’ multiple conformational states remains a critical research direction. Developing advanced computational strategies to accurately map out alternative conformations would significantly amplify the practical applicability of these models, especially in the field of drug design, where the ability to target specific protein states could lead to breakthrough interventions.

\textbf{Harnessing AlphaFold-latest and RoseTTAFold All-Atom:} Building upon AlphaFold's legacy, AlphaFold-latest opens new horizons in modeling complex molecular assemblies. Future research could be directed towards leveraging this model’s superior performance in protein-ligand and protein-nucleic acid interactions to facilitate breakthroughs in drug design and to gain a comprehensive understanding of biomolecular interactions. Although our initial plan included incorporating the latest developments from DeepMind - AlphaFold-latest \cite{GoogleDeepMindAlphaFoldTeamIsomorphicLabsTeam:2023} and RoseTTAFold All-Atom (RFAA) from the Institute for Protein Design at the University of Washington \cite{Krishna_etal_2023}, we faced constraints due to limited access to detailed technical information. Despite these limitations, our approach aims to surpass the existing barriers of current systems, particularly in enhancing DeepTracer's capacity to process and accurately predict structures from low-resolution cryo-EM maps.

%% file: tex/conclusion.tex
\section{Conclusions and Future Directions}

We propose DeepTracer-LowResEnhance, which combines established computational methodologies with innovative engineering approaches to significantly advance protein structure prediction. By leveraging systems like AlphaFold and introducing novel techniques, we have successfully extended the capability to analyze low-resolution cryo-EM maps. Our approach mitigates the limitations of existing map-to-model pipelines, expanding the applicability of DeepTracer to include maps with resolutions as low as 8.5 Å, where performance typically diminishes in conventional methods. This strategic enhancement underscores our methodology's effectiveness and broadens its usability across a wider resolution spectrum. The advancements presented have significant implications for structural biology, enabling more accurate and detailed analysis of low-resolution cryo-EM maps and helping to close the gap between available cryo-EM data and resolved protein structures.

In conclusion, our study not only elevates the current state of protein structure prediction but also charts a course for addressing the multifaceted challenges within structural biology. By embracing the limitations and potential avenues for advancement highlighted through our research, we are setting the stage for transformative developments in understanding and predicting the intricate world of biomolecular structures.